\documentstyle[12pt,epsfig]{article}
\begin{document}
\font\ninerm = cmr9
\def\footnoterule{\kern-3pt \hrule width
\hsize \kern2.5pt} \pagestyle{empty}
\vskip 0.5 cm

\begin{center}
{\large\bf IMPLICATIONS OF SPACETIME QUANTIZATION FOR\\
THE BAHCALL-WAXMAN NEUTRINO BOUND}
\end{center}
\vskip 0.5 cm

\begin{center}
{\bf Giovanni AMELINO-CAMELIA}$^a$,
{\bf Michele ARZANO}$^b$, {\bf Y. Jack NG}$^b$,\\
{\bf Tsvi PIRAN}$^c$
and {\bf Hendrik VAN DAM}$^b$\\
{\it $^a$Dipart.~Fisica, Univ.~Roma ``La Sapienza'' and INFN Sez.~Roma1\\
P.le Moro 2, 00185 Roma, Italy}\\
 {\it $^b$Institute of Field Physics,
 Department of Physics and Astronomy,\\
 University of North Carolina,
 Chapel Hill, NC 27599-3255, USA}\\
 {\it $^c$Racah Institute of Physics, Hebrew University,
Jerusalem 91904, Israel}
\end{center}

\vspace{1cm}
\begin{center}
{\bf ABSTRACT}
\end{center}

{\leftskip=0.6in \rightskip=0.6in There is growing interest in
quantum-spacetime models in which small  departures from Lorentz
symmetry are governed by the Planck scale. In particular, several
studies have considered the possibility that these small
violations of Lorentz symmetry may affect  various astrophysical
observations, such as the evaluation of the GZK limit for cosmic
rays, the interaction of TeV photons with the Far Infrared
Background and the arrival time of photons with different
energies from cosmological sources. We show that the same
Planck-scale departures from Lorentz symmetry that
lead to a modification of the GZK limit which would be consistent
with the observations reported by AGASA, also have significant
implications for the evaluation of the Bahcall-Waxman bound on
the flux of high-energy neutrinos produced by photo-meson
interactions in sources of size not much larger than the proton
photo-meson mean free path. }

%\pacs{{\tt$\backslash$\string pacs\{\}} }

\newpage
\baselineskip 12pt plus .5pt minus .5pt \pagenumbering{arabic}
 \pagestyle{plain}

\section{Introduction}
There is  a growing interest recently~[1-11]
%\cite{grbgac,garayPRL,gampul,thresh,gacTP,qgpMORE,qgprev,susskind,dsr1,jurekNEWdsrkmink,contraction}
in the possibility that some novel quantum properties of
spacetime may have important implications for the analysis of
Lorentz transformations. Some approaches to the quantum-gravity
problem attribute to the Planck scale $E_p$ ($E_p \sim 10^{28}eV$)
the status of an intrinsic characteristic of space-time
structure. For example $E_p$ can have a role in spacetime
discretization or in the commutation relations between spacetime
observables. It is very hard~\cite{dsr1} (perhaps even impossible)
to construct discretized versions or non-commutative versions of
Minkowski space-time which enjoy ordinary Lorentz symmetry.
Pedagogical illustrative examples of this observation have been
discussed, {\it e.g.}, in Ref.\cite{hooftlorentz} for the case of
discretization and in Refs.\cite{majrue,kpoinap,alessfranc} for
the case of non-commutativity. The action of ordinary (classical)
boosts on discretization length scales (or non-commutativity
length scales) will naturally be such that different inertial
observers would attribute different values to these lengths
scales, just as one would expect from the mechanism of
FitzGerald-Lorentz contraction.

Models based on an approximate Lorentz symmetry, with
Planck-scale-dependent departures from exact Lorentz symmetry,
have been recently considered in most quantum-gravity research
lines, including models based on ``spacetime foam"
pictures~\cite{grbgac,garayPRL}, ``loop quantum gravity"
models~\cite{gampul,contraction}, certain ``string theory"
scenarios~\cite{susskind}, and ``noncommutative
geometry~\cite{dsr1,jurekNEWdsrkmink}.

Interest in tests of modifications of Lorentz symmetry has also
increased recently as a result of the
realization~\cite{thresh,gacTP,colgla,ngthresh} that these
modifications of Lorentz symmetry provide one of the possible
solutions of the so-called ``cosmic-ray paradox". The spectrum of
observed cosmic rays was expected to be affected by a cutoff at
the scale $E_{\rm GZK} \sim 5 {\cdot} 10^{19} \; \hbox{eV}$.
Cosmic rays emitted with energy higher than $E_{\rm GZK}$ should
interact with photons in the cosmic microwave background and lose
energy by pion emission, so that their energy should have been
reduced to the $E_{\rm GZK}$ level by the time they reach our
Earth observatories. However, the AGASA observatory has
reported~\cite{agasa} several observations of cosmic rays with
energies exceeding the $E_{\rm GZK}$ limit~\cite{GZK} by nearly
an order of magnitude. As other experiments do not see an excess
of particles above the GZK limit,  this experimental puzzle will
only be established when confirmed by larger observatories, such
as Auger~\cite{auger}. Furthermore, numerous other solutions have
been discussed in the literature. Still,  it is noteworthy that
Planck-scale modifications of Lorentz symmetry can
raise~\cite{thresh,gacTP} the threshold energy for pion
production in collisions between cosmic rays and microwave
photons, and the increase is sufficient to explain away the
puzzle associated with the mentioned ultra-high-energy cosmic-ray
observations.

Bahcall-Waxman~\cite{bw} have shown that the same particles which
we observe as high-energy cosmic rays should also lead to neutrino
production at the source. Using the observed cosmic ray fluxes
they derive a bound (the Bahcall-Waxman bound) on the  flux of
high-energy neutrinos that can be revealed in astrophysics
observatories. We show here  that the same Lorentz-symmetry
violations that can extend the cosmic-ray spectrum also affect
the chain of particle-physics processes that arise in the
neutrino production and hence in  establishing the Bahcall-Waxman
bound~\cite{bw}. Thus, the departures from Lorentz symmetry that
are capable of explaining the ``cosmic-ray paradox" inevitably
lead to modification of this limit.

Since a relatively large variety of quantum-gravity pictures is
being considered in the analysis of the cosmic-ray paradox, in
this first study we only intend to illustrate our point within a
simple phenomenological model, leaving for future studies the
task of more precise analysis of specific quantum-gravity models.
For similar reasons we only focus on one of the chains of
processes that are relevant for the Bahcall-Waxman bound: the
case in which a proton at the source undergoes photo-pion
interactions of the type $p + \gamma \rightarrow X + \pi^+$
before escaping the source, then giving rise to neutrino
production through the decays $\pi^+ \rightarrow \mu^+ + \nu_\mu$
and $ \mu^+ \rightarrow  e^+ + \nu_e + {\bar{\nu}}_\mu$.

The phenomenological model
on which we focus is the simplest one in the literature, which
evolved primarily through the studies reported in
Refs.~\cite{grbgac,thresh,gacTP}. This is a kinematic in which
the Planck-scale $E_p$ enters the energy/momentum dispersion
relation
\begin{equation}
m^2 = E^2 - \vec{p}^2 + f(E,\vec{p};E_p) \simeq
E^2 - \vec{p}^2 + \eta \vec{p}^2 \left({E \over E_p}\right)^n
%\vec{p}^2
~
\label{displead}
\end{equation}
while the laws of energy-momentum conservation remain
unaffected\footnote{Our analysis
based on
(\ref{displead}) and standard energy-momentum conservation
should be applicable (up to small
numerical modifications) to a large
class of quantum-gravity models which are being considered as
possible solutions of the cosmic-ray paradox. One exception is
the ``doubly-special relativity" framework \cite{dsr1}, in which one
could adopt a dispersion relation of type (\ref{displead}) but it
would then be necessary to introduce a corresponding modification
of the laws of energy-momentum conservation (in order to avoid the
emergence of a preferred class of inertial observers~\cite{dsr1}).
Our conclusions are not applicable to that scheme.} by the Planck
scale. $\eta$ is a dimensionless coefficient which one expects to
be roughly of order $1$ (but cannot be reliably predicted at the
present preliminary level of development of the relevant
quantum-gravity models). The power $n$, which should also be
treated as a phenomenological parameter, is a key element of this
phenomelogical scenario, since it characterizes the first
nonvanishing contribution in a (inverse-)Planck-scale power
series of the quantum-gravity-induced correction
$f(E,\vec{p};E_p)$. It is usually expected that $n = 1$ and $n =
2$ are most likely.

In Section~2 we revisit the analysis of the emergence
of ``threshold anomalies" due to (\ref{displead}) in the study of
particle production in collision processes. For positive $\eta$
($\eta \sim 1$) and $n \le 2$, according to the Planck-scale
effect (\ref{displead}) one expects~\cite{gacTP} an increase in
the threshold energy for pion production in collisions between
cosmic rays and microwave photons, and the increase is sufficient
to explain away the GZK puzzle raised by the AGASA observations.
We also comment on another potentially observable threshold
anomaly that concerns electron-positron pair production in
photon-photon collisions. Throughout Section~2 we also emphasize
the differences between the case of positive $\eta$ and the case
of negative $\eta$.

In Section~3 we show that (\ref{displead}) also affects
significantly the at-the-source processes of the type $p + \gamma
\rightarrow X + \pi^+$ that are relevant for the Bahcall-Waxman
analysis. If $\eta \sim 1$ and $n \le 2$ ({\it i.e.} for the same
departures from Lorentz symmetry that would explain the
cosmic-ray paradox) (\ref{displead}) leads to the prediction of a
strongly reduced probability for the process $p + \gamma
\rightarrow X + \pi^+$ to occur before the proton escapes the
source. Correspondingly one expects sharply reduced neutrino
production, and as a result the ``quantum-gravity-modified
Bahcall-Waxman bound" should be expected to be many orders of
magnitude lower than the standard Bahcall-Waxman bound. The
opposite effect is found for negative $\eta$ ($\eta \sim - 1$):
in that case one would expect the standard Bahcall-Waxman bound
to be violated, {\it i.e.} for negative $\eta$ one could find a
neutrino flux that exceeds the  standard Bahcall-Waxman bound.

In Section~4 we show that also the decays $\pi^+ \rightarrow
\mu^+ + \nu_\mu$ and $ \mu^+ \rightarrow  e^+ + \nu_e +
{\bar{\nu}}_\mu$ are significantly affected by the Planck-scale
effect (\ref{displead}). Again the effect goes in the direction
of reducing neutrino production for positive $\eta$. However we
also observe that the dominant quantum-gravity modification of
the Bahcall-Waxman bound comes at the level of analysis of
processes of the type $p + \gamma \rightarrow X + \pi^+$, where a
several-order-of-magnitude modification would be expected,
whereas the additional modification encountered at the level of
the processes $\pi^+ \rightarrow \mu^+ + \nu_\mu$ and $ \mu^+
\rightarrow  e^+ + \nu_e + {\bar{\nu}}_\mu$ is not as significant.
Section~5 is devoted to our closing remarks.

\section{Previous results on Planck-scale-induced threshold anomalies and the
sign of $\eta$}

We begin by considering the implications of Eq.~(\ref{displead})
for the analysis of processes of the type $ 1 + \gamma
\rightarrow 2 + 3$. The key point for us is that, for a given
energy $E_1$ of the particle that collides with the photon, there
is of course a minimal energy $\epsilon_{min}$ of the photon
($\gamma$) in order for the process to be kinematically allowed,
and therefore achieve production of the particles $2$ and $3$.
One finds that if $\eta$ is positive the value of
$\epsilon_{min}$ predicted according to the Planck-scale effect
(\ref{displead}) is higher than the corresponding value obtained
using ordinary Lorentz symmetry.

In the applications that are of interest here the particle that
collides with the photon has a
very high energy, $E_1 \simeq p_1 \gg m_1$,
and its energy is also much larger than the energy of
the photon with which it collides $E_1 \gg \epsilon$. This will
allow some useful simplifications in the analysis.

Let us start by briefly summarizing the familiar derivation of
$\epsilon_{min}$ in the ordinary Lorentz-invariant case.  At the
threshold (no  momenta in the CM frame after the collision)
energy conservation and momentum conservation become one
dimensional:
\begin{equation}
E_1+\epsilon=E_2+E_3 ~, \label{econsv}
\end{equation}
\begin{equation}
p_1-q=p_2+p_3~, \label{pconsv}
\end{equation}
where $q$ is the photon's momentum. The ordinary
Lorentz-invariant relations are
\begin{equation}
q=\epsilon~,~~~E_i = \sqrt{p_i^2+m_i^2}
\simeq p_i + {m_i^2 \over 2 p_i}
~,
\label{lirel}
\end{equation}
where we have assumed that, since $E_1$ is large as mentioned,
$E_2$ and $E_3$ are also large ($E_{2,3} \simeq p_{2,3} \gg m_{2,3}$).

The threshold conditions are usually identified by transforming
these laboratory-frame relations into center-of-mass-frame
relations and imposing that the center-of-mass energy be equal to
$m_2+m_3$.  However, in preparation for the discussion of
deformations of Lorentz invariance it is useful to work fully in
the context of the laboratory frame. There the threshold condition
that characterizes $\epsilon_{min}$ can be identified with the
requirement that the solutions for $E_2$ and $E_3$ as functions
of $\epsilon$ (with a given value of $E_1$) that follow from
Eqs.~(\ref{econsv}), (\ref{pconsv}) and (\ref{lirel}) should be
imaginary for $\epsilon < \epsilon_{min}$ and should be real for
$\epsilon \ge \epsilon_{min}$. This straightforwardly leads to
\begin{equation}
\epsilon \ge \epsilon_{min} \simeq {(m_2 + m_3)^2 - m_1^2 \over 4 E_1}
~.
\label{lithresh}
\end{equation}

This standard Lorentz-invariant analysis is
modified~\cite{thresh,gacTP} by the deformations codified in
(\ref{displead}).  The key point is that Eq.~(\ref{lirel}) is
replaced by
\begin{equation}
\epsilon= q - \eta {q^{n+1} \over 2 E_p^n} ~,~~~ E_i \simeq p_i +
{m_i^2 \over 2 p_i}- \eta {p_i^{n+1} \over 2 E_p^n} ~.
\label{lv1rel}
\end{equation}
Combining (\ref{econsv}), (\ref{pconsv}) and (\ref{lv1rel}) one
obtains a modified kinematical requirement
\begin{equation}
\epsilon \ge \epsilon_{min} \simeq  {(m_2 + m_3)^2 - m_1^2 \over 4 E_1}
+ \eta {E_{1}^{n+1} \over 4 E_p^n} \left( 1 -
{m_2^{n+1} + m_3^{n+1} \over (m_2 + m_3)^{n+1}}  \right)
~.
\label{lithresh2}
\end{equation}
where we have included only the leading corrections (terms suppressed
by both the smallness of $E_{p}^{-1}$ and the smallness of $\epsilon$
or $m$ were neglected).

The Planck-scale ``threshold anomaly"~\cite{gacTP} described by
Eq.~(\ref{lithresh2}) is relevant for the analysis of the GZK
limit in cosmic-ray physics. In fact, the GZK limit essentially
corresponds to the maximum energy allowed of a proton in order to
travel in the CMBR without undergoing processes of the type $p +
\gamma \rightarrow p + \pi$. For a proton of energy $E_1 \sim 5
{\cdot} 10^{19}eV$ the value of $\epsilon_{min}$ obtained from
the undeformed equation (\ref{lithresh}) is such that CMBR
photons can effectively act as targets for photopion production.
But, for $\eta \sim 1$ and $n \le 2$, the value of
$\epsilon_{min}$ obtained from the Planck-scale deformed equation
(\ref{lithresh2}) places CMBR photons below threshold for
photopion production by protons with energies as high as $E_1
\sim 10^{21}eV$, and would explain~\cite{thresh,gacTP} observations
of cosmic rays above the GZK limit. For negative $\eta \sim - 1$
one obtains the opposite result: photopion production should be
even more efficient than in the standard case. Therefore negative
$\eta$ is disfavoured by the observations reported by various
UHECR observations.

There has also been some interest~\cite{thresh,gacTP,qgpMORE} in
the implications of Eq.~(\ref{lithresh2}) for electron-positron
pair production in collisions between astrophysical high-energy
photons and the photons of the Far Infrared Background.
Electron-positron pair production should start to be significant
when the high-energy photon has energies of about 10 or 20 TeV.
The Planck-scale correction in Eq.~(\ref{lithresh2}) would be
significant, though not dominant, at those energies. Observations
of TeV photons are becoming more abundant, but the field is still
relatively young.  Moreover, our knowledge of the Far Infrared
Background is presently not as good as our knowledge of the CMBR.
Therefore observations of TeV photons do not yet provide a
significant insight on the Planck scale physics of interest here.
Consistency with those observations only imposes a
constraint~\cite{qgpMORE} of the type $|\eta| < 100$, which
(since the quantum-gravity intuition favours $|\eta| \sim 1$) is
not yet significant from a quantum-gravity perspective.

A similar upper limit ($|\eta| < 100$) is obtained by considering
the implications~\cite{grbgac,billetal} of the deformed dispersion
relation for the arrival times of  photons with different energies
emitted (nearly-)simultaneously from cosmological sources.

In summary, the present situation justifies some interest for the
case of positive $\eta$, particularly as a possible description
of cosmic rays above the GZK limit. The case of negative $\eta$
is disfavored by various UHECR observations. Additional
phenomenological reasons to disfavour negative $\eta$ have been
found in analyses of photon stability (see, {\it e.g.},
Ref.~\cite{qgpMORE}), which is instead not relevant for the
positive $\eta$ case. Moreover, the case of negative $\eta$
appears to be also troublesome conceptually since it leads to
superluminal velocities in a framework, such as the one adopted
here, in which the new effects are simply motivated by the idea
of a quantum-spacetime medium\footnote{A deformed dispersion
relation is generically expected in a special relativistic theory
when a medium is present. The presence of the medium does not
alter the principles of special relativity, and superluminal
velocities should not be allowed. The situation is different in
the context of the approach proposed in Ref.~\cite{dsr1}, in
which the deformed dispersion relation is not motivated by the
presence of a quantum-spacetime medium but rather by a role for
the Planck scale in the relativity principles. In the framework of
Ref.~\cite{dsr1} superluminal velocities would not lead to
paradoxical results.}, and therefore do not naturally lead to the
expectation of superluminal velocities. Still, as a contribution
to this evolving understanding, we will consider the
Bahcall-Waxman bound both for positive and negative $\eta$.

\section{Planck-scale-induced threshold anomalies and the
neutrino bound} A key observation for our analysis comes from the
fact that the Planck-scale threshold anomaly described by
Eq.~(\ref{lithresh2}) is significant for the Bahcall-Waxman bound
for the same reasons that render it significant for the GZK limit
in cosmic-ray physics. In fact, both the Bahcall-Waxman bound and
the GZK limit involve the analysis of processes of the type $p +
\gamma \rightarrow X + \pi$ in which a high-energy proton
collides with a softer photon. In the case of the Bahcall-Waxman
bound one finds that for a proton of energy $E_1 \sim 10^{19}eV$
which is emerging from a source ({\it e.g.} an AGN), according to
the standard kinematical requirement (\ref{lithresh}) the photons
in the environment that are eligible for production of charged
pions $\pi^+$ are all the photons with energy $\epsilon \ge
\epsilon_{min} \sim 0.01 eV$. But, for the Planck-scale scenario
of (\ref{lithresh2}) with $\eta \sim 1$ and $n = 1$ far fewer
photons in the environment, viz. only photons with energy
$\epsilon \ge \epsilon_{min} \sim 10^9 eV$, are kinematically
eligible for production of charged pions. In a typical source the
abundance of photons with  $\epsilon \ge  10^9 eV$ is much
smaller, by several orders of magnitude, than the abundance of
photons with $\epsilon \ge 0.01 eV$. Correspondingly the
Planck-scale effect predicts a huge reduction in the probability
that a charged pion be produced before the proton escapes the
source, and in turn this leads (for $\eta >0$ and $n=1$) to a
decrease in the expected high-energy neutrinos flux by many
orders of magnitude below the level set by the Bahcall-Waxman
bound.

The same qualitative picture applies to the
case $\eta \sim 1$, $n = 2$, although the effect is somewhat
less dramatic because of the large suppression of the
effect that is due to the extra power of the Planck scale.
In fact, for $\eta \sim 1$, $n = 2$ one finds that
the photons in the environment that are energetically enough for the
production of charged pions must have
energy $\epsilon \ge \epsilon_{min} \sim 1 eV$.

Whereas for positive $\eta$ the Planck-scale effect leads
to a lower neutrino bound the reverse is true for negative $\eta$.
In particular, for $\eta \sim - 1$, $n \le 2$
from (\ref{lithresh2}) it follows that photons in the source
with energies even below\footnote{Formally in this case
Eq.~(\ref{lithresh2}) even admits photon targets with ``negative energies".
But, of course, considering the approximations we implemented,
one can only robustly infer that photons with very low energies
can lead to pion production.} $0.01 eV$
are viable targets for the production of charged pions
by protons with energy $E_1 \sim 10^{19}eV$.
Correspondingly, the Bahcall-Waxman bound would be weakened.

\section{Implications of the Planck-scale for particle decays
and the neutrino bound}

In the previous Section we have shown that the Planck-scale
effects considered here would affect the production of charged
pions before the ultra-high-energy cosmic-ray proton escapes the
source. In this Section we analyze the implications of the same
effects for the decay processes $\pi^+ \rightarrow \mu^+ +
\nu_\mu$ and  $ \mu^+ \rightarrow  e^+ + \nu_e + {\bar{\nu}}_\mu$
which are also relevant for the Bahcall-Waxman bound.

Since we are interested in both a two-body
decay, $\pi^+ \rightarrow \mu^+ + \nu_\mu$,
and a three-body
decay, $\mu^+ \rightarrow  e^+ + \nu_e + {\bar{\nu}}_\mu$
it is convenient for us to obtain a general result for $N$-body
decays.
This will also be a technical contribution to the study
of the kinematics governed by (\ref{displead}).
In fact, the implications of (\ref{displead}) for two-body decays
have been previously analyzed~\cite{gacpion},
but for decays in three or more particles
there are no previous results in the literature.

We start our analysis of the decay $A \rightarrow 1 + 2 + ... +N$
($A$ is the generic particle that decays into
particles $1,  2,  ... N$)
with the energy-momentum conservation laws:
\begin{equation}
\label{cons1}
E_A=E_1+E_2+...+E_N
\end{equation}
\begin{equation}\label{cons2}
\vec{p}_A=\vec{p}_1+\vec{p}_2+...+\vec{p}_N
\end{equation}

Denoting with $\theta_{ij}$ the angle between the linear
momentum of particle $i$ and that of particle $j$,
and denoting with $p$ the modulus of the 3-vector $\vec{p}$,
we can use (\ref{cons2}) to obtain ($i,j=1,2,...N$)
\begin{equation}\label{sq2}
p_A^2=\sum_ip_i^2+\sum_{i\neq j}p_ip_j\cos \theta_{ij}
\end{equation}
and (\ref{cons1}) to get
\begin{equation}\label{sq1}
E_A^2=\sum_iE_i^2+\sum_{i\neq j}E_iE_j\,\, .
\end{equation}

From (\ref{sq2}) and (\ref{sq1})
it follows that
\begin{equation}\label{diff1}
E_A^2-p_A^2=\sum_i(E_i^2-p_i^2)+\sum_{i\neq j}(E_iE_j-p_ip_j\cos \theta_{ij})\,\, .
\end{equation}
Next we use the deformed dispersion
relation (\ref{displead}), $E^2-p^2=m^2-\eta E^n p^2/E_p^n$,
to obtain
\begin{equation}\label{diff2}
m_A^2-\eta E_A^2 p_A^2/E_p^2
=\sum_i(m^2_i-\eta E_i^n p_i^2/E_p^n)
+\sum_{i\neq j}(E_iE_j-p_ip_j\cos \theta_{ij})
\end{equation}

For simplicity let us consider separately the cases $n=1$ and $n=2$,
starting with $n=1$.
It is convenient to rewrite the kinematical condition (\ref{diff2}),
for $n=1$, in the following way
\begin{equation}\label{diff4}
\sum_im^2_i-m_A^2+\sum_{i\neq j}\left(p_ip_j+p_i\frac{m^2_j}{p_j}\right)
+ {\eta \over E_p}\left(E_A^3-\sum_i E_i^3-\sum_{i\neq j}E_i E_j^2\right)
= \sum_{i\neq j}p_ip_j\cos \theta_{ij}
\end{equation}
where we used again the deformed dispersion relation,
\begin{equation}\label{Eexp}
E=(p^2+m^2-{\eta \over E_p} E p^2)^{\frac{1}{2}}\simeq p
-\frac{\eta}{2E_p}p^2+\frac{m^2}{2p} ~.
\end{equation}
We are neglecting terms of order $E_p^{-2}$ and higher, which are
clearly subleading, and we are also neglecting terms of
order $E_p^{-1} m^2$ which are negligible compared to terms of
order $E_p^{-1} E^2$ since all particles involved in the processes
of interest to us have very high momentum.

Using the fact that $\cos \theta_{ij} \le 1$ for every $\theta_{ij}$,
it follows
from (\ref{diff4}) that for the
decay to be kinematically allowed a necessary condition is
\begin{equation}\label{cond1}
\sum_im^2_i-m_A^2+\sum_{i\neq j}\left(p_ip_j+p_i\frac{m^2_j}{p_j}\right)
+{\eta \over E_p}\left(E_A^3-\sum_i E_i^3-\sum_{i\neq j}E_i E_j^2\right)
\leq\sum_{i\neq j}p_ip_j
\end{equation}
or equivalently
\begin{equation}\label{cond2}
\sum_im^2_i-m_A^2+\sum_{i\neq j}p_i\frac{m^2_j}{p_j}
+{\eta \over E_p}\left(E_A^3-\sum_i E_i^3-\sum_{i\neq j}E_i E_j^2\right)
\leq 0
\end{equation}

In the analysis of particle-decay processes relations of the type
(\ref{cond2}) impose constraints on the available phase space.
For positive $\eta$ the
quantum-gravity effect clearly goes in the direction of
reducing the available phase space; in fact, it is easily seen
that
\begin{equation}
\left(E_A^3-\sum_iE_i^3-\sum_{i\neq j}E_iE_j^2\right)
=\left((\sum_iE_i)^3-\sum_iE_i^3-\sum_{i\neq j}E_iE_j^2\right)> 0
~.
\end{equation}
The correction is completely negligible as long as $m_A^2 \gg
E_A^3/E_p$, but for $m_A^2 \ll E_A^3/E_p$ there is clearly a
portion of phase space in which, for positive $\eta$, condition
(\ref{cond2}) is not satisfied. (Think for example of the case
$E_1 \sim E_2 \sim ... \sim E_N \sim E_A/N$.) Starting at $E_A
\ge (m_A^2 E_p)^{1/3}$ the phase space available for the decay of
particle $A$ is gradually reduced as $E_A$ increases. The
difference between the standard Lorentz-symmetry prediction for
the lifetime and the quantum-gravity-corrected prediction becomes
more and more significant as the energy of the decaying particle
is increased, and goes in the direction of rendering the particle
more stable, i.e., rendering the decay more unlikely. For the
relevant decays of pions and muons we expect that the
quantum-gravity effect starts being important at pion/muon
energies of order $(m_\pi^2 E_p)^{1/3} \sim (m_\mu^2 E_p)^{1/3}
\sim  10^{15} eV$.

For positive $\eta$ it is inevitable that at some energy a
significant suppression of the decay probability kicks in, while
for negative $\eta$ it is easy to see that there is no effect on
the size of the phase space  available for the decays. The change
in the sign of $\eta$ turns as usual into a change of sign of the
effect, which would go in the direction of extending the phase
space available for the decay, but the relevant portion of
parameter space (some neighborhood of $E_1 \sim E_2 \sim ... \sim
E_N \sim E_A/N$) is already allowed even without the Planck scale
effect, so for negative-$\eta$ effect is not significant.

Completely analogous considerations apply to the case $n=2$.
From (\ref{diff2}),
for $n=2$, one obtains
\begin{equation}
\sum_im^2_i-m_A^2+\sum_{i\neq j}\left(p_ip_j+p_i\frac{m^2_j}{p_j}\right)
+ {\eta \over E_p^2}\left(E_A^4-\sum_i E_i^4-\sum_{i\neq j}E_i E_j^3\right)
= \sum_{i\neq j}p_ip_j\cos \theta_{ij}
~,
\end{equation}
and then, just following the same line of analysis we already adopted
for the case $n=1$, for $n=2$ one finds that the decay is only
allowed if
\begin{equation}
\sum_im^2_i-m_A^2+\sum_{i\neq j}p_i\frac{m^2_j}{p_j}
+{\eta \over E_p^2}\left(E_A^4-\sum_i E_i^4-\sum_{i\neq j}E_i E_j^3\right)
\leq 0
~.
\end{equation}
Since
\begin{equation}
\left(E_A^4-\sum_iE_i^4-\sum_{i\neq j}E_iE_j^3\right)
=\left((\sum_iE_i)^4-\sum_iE_i^4-\sum_{i\neq j}E_iE_j^3\right)> 0
~,
\end{equation}
we find again that for positive $\eta$
the quantum-gravity correction inevitably goes
in the direction of reducing the available phase space and therefore
rendering the decay more unlikely.
In this $n=2$ case we expect that the quantum-gravity suppression
of the decay probability starts being important at pion/muon
energies of
order $(m_\pi E_p)^{1/2} \sim (m_\mu E_p)^{1/2} \sim 10^{18} eV$.

Of course also for $n=2$ one finds that the case of negative $\eta$
does not have significant implications, for exactly the same
reasons discussed above in considering the $n=1$ case.

\section{Closing remarks}
We found that Planck-scale effects can have important implications
for the neutrino-producing chain of
processes $p + \gamma \rightarrow X + \pi^+$,
 $\pi^+ \rightarrow \mu^+ + \nu_\mu$,
 $\mu^+ \rightarrow  e^+ + \nu_e + {\bar{\nu}}_\mu$, which are
relevant for the Bahcall-Waxman bound. We focused on a single
simple example of Planck-scale kinematics. Since this analysis
led to encouraging conclusions (significant implications for the
Bahcall-Waxman bound) it should provide motivation for more
detailed studies in more structured quantum-gravity models.
Because of the simple kinematical origin of our argument it is
reasonable to expect that these more detailed studies will
confirm that, for positive $\eta$, the quantum-gravity effect
leads to a neutrino flux that is many orders of magnitude below
the level allowed by the Bahcall-Waxman bound. We have shown here
that this is due primarily to a strong suppression of the
production of high-energy charged pions by protons at the source.
If future observations give us a neutrino flux which is close to
the level allowed by the Bahcall-Waxman bound, the type of
quantum-gravity physics considered here would be excluded (for
positive $\eta$). On the other hand, a low neutrino flux will be
harder to interpret as, a priori, it is not clear that the
Bahcall-Waxman bound should be saturated.

The suppression present for positive $\eta$ already found
full support at the first step in the chain of processes,
in the collisions $p + \gamma \rightarrow X + \pi^+$.
We felt however that it was appropriate to consider also the processes
further down in the chain, in the
decays of $\pi^+ \rightarrow \mu^+ + \nu_\mu$,
$ \mu^+ \rightarrow  e^+ + \nu_e + {\bar{\nu}}_\mu$.
In fact, it was conceivable that perhaps at that level the production
of neutrinos might receive a compensating boost form the quantum-gravity
effect which, for positive $\eta$, suppresses the likelihood
of the process $p + \gamma \rightarrow X + \pi^+$.
This turned out not to be the case: for positive $\eta$ one actually
expects a further suppression of neutrino production, since the
quantum-gravity effect renders ultrahigh-energy pions and muons
more stable.
This part of our analysis also provided a technical contribution to the
study of the kinematics governed by (\ref{displead}), since
previously the implications of (\ref{displead}) were only known
for two-body decays, while here we  obtained a generalization
to $N$-body decays for arbitrary $N$.

As discussed in Section~2, the case of negative $\eta$ is
disfavoured conceptually and starts to be strongly constrained by
preliminary observations in astrophysics. We found that it would
have striking consequences for the Bahcall-Waxman bound:
in the case of negative $\eta$ the
modification of the Bahcall-Waxman bound would amount to violating
(raising) the Bahcall-Waxman bound by several orders of magnitude.

Our analysis contributes to ongoing work aimed at establishing a
web of consequences of the type of Planck-scale kinematics
considered here. It is not hard to find several different
solutions to a single anomaly in ultrahigh-energy astrophysics,
{\it e.g.} the cosmic-ray paradox, if confirmed by other
observatories. However for the type of Planck-scale kinematics
considered here, there are several correlated predictions and
these together can be used to favor or rule out the scenario. In
particular, evidence supporting both a cosmic-ray paradox and an
unexpectedly low ultrahigh-energy neutrino flux would fit
naturally within the Planck-scale-kinematics scenario (with
positive $\eta$). More precisely, evidence supporting both a
cosmic-ray paradox and an unexpectedly low ultrahigh-energy
neutrino flux would favor solutions of the cosmic-ray paradox
based on violations of Lorentz symmetry with respect to other
proposed solutions of the cosmic-ray paradox. In fact, the
correlation we have exposed here between a cosmic-ray paradox and
a lowered Bahcall-Waxman bound is a characteristic models in
which the kinematics of the processes $p + \gamma \rightarrow X +
\pi$ is modified by a violation of Lorentz symmetry. In fact, the
processes $p + \gamma \rightarrow X + \pi^{{\pm},0}$ dominate the
GZK threshold and the Bahcall-Waxman limit. An increase in the
GZK limit and a lowered Bahcall-Waxman bound are found whenever
the violation of Lorentz symmetry causes an increased
energy-threshold condition for the processes $p + \gamma
\rightarrow X + \pi$. Therefore it can distinguish between models
with and without this Lorentz-violation effect, but it cannot
establish whether the origin of the Lorentz violation is
connected with quantum gravity. In particular, it is interesting
to consider the suggestion of Coleman and Glashow~\cite{colgla}
concerning a specific Lorentz-violation solution of the GZK
paradox, which is not motivated by quantum gravity. Coleman and
Glashow~\cite{colgla} consider a scheme in which different
particles have a different ``maximum attainable speed"
(essentially a different ``speed-of-light constant" for different
particles). This can be cast into our formalism with a
particle-dependent $\eta$ and with $n=0$. It follows from our
analysis that any model that resolves the UHECR GZK paradox using
the Coleman-Glashow scheme will also lead to a stronger
Bahcall-Waxman neutrino bound.

\section*{Acknowledgments}
This work was supported
in part by the US Department of Energy and the Bahnson Fund of the
University of North Carolina.


\begin{thebibliography}{99}

\bibitem{grbgac} G.~Amelino-Camelia, J.~Ellis, N.E.~Mavromatos,
D.V.~Nanopoulos and S.~Sarkar,
% {\it Tests of quantum gravity from observations
% of $\gamma$-ray bursts},
astro-ph/9712103,
%Nature {393} (1998) 763-765.
Nature {393}, 763 (1998).

\bibitem{garayPRL} L.J.~Garay,
Phys.~Rev.~Lett.~{80} (1998) 2508.

\bibitem{gampul} R.~Gambini and J.~Pullin,
%{\it Nonstandard optics from quantum spacetime},
gr-qc/9809038,
Phys.~Rev.~D59 (1999) 124021; J.~Alfaro,
H.A.~Morales-Tecotl and L.F.~Urrutia,
%{\it Loop quantum gravity and light propagation},
%hep-th/0108061, Phys.~Rev.~D65 (2002) 103509;
%{\it Quantum Gravity corrections to neutrino propagation},
gr-qc/9909079,
Phys.~Rev.~Lett.~84 (2000) 2318.

\bibitem{thresh} T.~Kifune,
%{\it Invariance violation extends the cosmic ray horizon?},
Astrophys.~J.~Lett.~{518}, L21 (1999);
R.~Aloisio, P.~Blasi, P.L.~Ghia and A.F.~Grillo,
%{\it Probing The Structure of Space-Time with Cosmic Rays},
astro-ph/0001258,
Phys.~Rev.~D62 (2000) 053010;
R.J.~Protheroe and H.~Meyer,
%{\it An infrared background TeV gamma ray crisis?},
Phys.~Lett.~{B493} (2000) 1.

\bibitem{gacTP}  G. Amelino-Camelia and T. Piran,
astro-ph/0008107,
Phys.~Rev.~{D64} (2001) 036005.

\bibitem{qgpMORE} T.~Jacobson, S.~Liberati and D.~Mattingly,
%{\it TeV Astrophysics Constraints on Planck Scale Lorentz Violation},
hep-ph/0112207;
T.J.~Konopka, S.A.~Major,
%{\it Observational Limits on Quantum Geometry Effects},
hep-ph/0201184, New J.~Phys.~4 (2002) 57;
G. Amelino-Camelia, gr-qc/0212002,
R.C.~Myers and M.~Pospelov,
%{\it  Ultraviolet modifications of dispersion relations in effective field
%theory}
hep-ph/0301124.

\bibitem{qgprev} G.~Amelino-Camelia, gr-qc/991089,
%{\it Are we at the dawn of quantum-gravity phenomenology?},
Lect.~Notes Phys.~541 (2000) 1;
N.E.~Mavromatos, gr-qc/0009045;
S.~Sarkar,
%{\it Possible Astrophysical Probes of Quantum Gravity},
%Invited talk at 1st IUCAA Workshop on Interface of Gravitational
%and Quantum Realms, Pune, India, 17-21 Dec 2001.
gr-qc/0204092;
D.V.~Ahluwalia,
%{\it Interface of Gravitational and Quantum realms},
%Invited talk at 1st IUCAA Workshop on Interface of Gravitational
%and Quantum Realms, Pune, India, 17-21 Dec 2001.
gr-qc/0205121;
Y.J.~Ng
%     Title: Selected topics in Planck-scale physics
gr-qc/0305019.

\bibitem{susskind}  A.~Matusis, L.~Susskind and N.~Toumbas,
%The IR/UV connection in the non-commutative gauge theories,
hep-th/0002075, JHEP 0012 (2000) 002.

\bibitem{dsr1} G.~Amelino-Camelia, gr-qc/0012051,
%{\it Relativity in space-times with short-distance
%structure governed by an observer-independent (Planckian)
%length scale},
%Int.~J.~Mod.~Phys.~{\bf D11}, 35 (2002) 35-60;
Int.~J.~Mod.~Phys.~{D11} (2002) 35;
hep-th/0012238,
%{\it Testable scenario for Relativity with minimum length},
%Phys.~Lett.~{\bf B510} (2001) 255-263.
Phys.~Lett.~{B510} (2001) 255.

\bibitem{jurekNEWdsrkmink} J.~Kowalski-Glikman and S.~Nowak,
%{\it Noncommutative space-time of doubly special relativity theories},
hep-th/0204245,
Int.~J.~Mod.~Phys.~D12 (2003) 299.

\bibitem{contraction} G.~Amelino-Camelia, L.~Smolin and A.~Starodubtsev,
hep-th/0306134.
%Quantum symmetry, the cosmological constant and Planck scale phenomenology

\bibitem{hooftlorentz}
G.~`t Hooft,
%Quantization of point particles in (2+1)-dimensional gravity and
%spacetime discreteness.
{Class.~Quant.~Grav.}~{\bf 13}, 1023 (1996).
%Class.~Quant.~Grav. {\bf 13} 1023--1039 (1996).

\bibitem{majrue} S.~Majid and H.~Ruegg,
%{\it Bicrossproduct structure of kappa {P}oincar{\'{e}} group
%and noncommutative geometry},
{Phys.~Lett.}~{B334} (1994) 348.

\bibitem{kpoinap} J.~Lukierski, H.~Ruegg and W.J.~Zakrzewski
%{\it Classical and quantum-mechanics
%of free $\kappa$-relativistic systems},
%Ann. Phys. {\bf 243} (1995) 90-116.
{Ann.~Phys.}~{243} (1995) 90.
%J.~Lukierski, A.~Nowicki, H.~Ruegg, and V.N. Tolstoy,
%{\it {$q$}-{D}eformation of {P}oincar{\'{e}} algebra},
%%%Phys. Lett. B264 (1991) 331-338;
%Phys. Lett. B264 (1991) 331;

\bibitem{alessfranc} A.~Agostini, G.~Amelino-Camelia, F.~D'Andrea,
%HOPF ALGEBRA DESCRIPTION OF NONCOMMUTATIVE SPACE-TIME SYMMETRIES.
hep-th/0306013.


\bibitem{colgla} S.~Coleman, S.L.~Glashow,
%{\it High-energy tests of Lorentz invariance},
Phys.~Rev.~D59 (1999) 116008.

\bibitem{ngthresh}
Y.J.~Ng, D.S.~Lee, M.C.~Oh, and H.~van Dam, Phys.~Lett.~B507, 236 (2001);
G. Amelino-Camelia, Y.J. Ng, and H. van Dam, gr-qc/0204077, Astropart.
Phys. (in press).

\bibitem{agasa} M.~Takeda et al., Phys. Rev. Lett.~81 (1998) 1163.

\bibitem{GZK} K.~Greisen, Phys.~Rev.~Lett.~16, 748 (1966);
G.~T.~Zatsepin and V.~A. Kuzmin, Sov.~Phys.-JETP Lett.~4, 78 (1966).

\bibitem{auger} Information on the Pierre Auger Observatory
is available at http://www.auger.org

\bibitem{bw} E.~Waxman and J.N.~Bahcall,
%HIGH-ENERGY NEUTRINOS FROM ASTROPHYSICAL SOURCES: AN UPPER BOUND.
hep-ph/9807282, Phys.~Rev.~D59 (1999) 023002;
J.N.~Bahcall and E.~Waxman,
%HIGH-ENERGY ASTROPHYSICAL NEUTRINOS: THE UPPER BOUND IS ROBUST.
hep-ph/9902383, Phys.~Rev.~D64 (2001) 023002.

\bibitem{billetal} S.D. Biller {\it et al},
%{\it Limits to Quantum Gravity Effects from Observations
%of TeV Flares in Active Galaxies},
Phys.~Rev.~Lett.~83 (1999) 2108.

\bibitem{gacpion} G. Amelino-Camelia, gr-qc/0107086,
Phys.~Lett.~{B528} (2002) 181.
%{\it Phys.~Lett.}~B {\bf 528}, 181-187 (2002).



\end{thebibliography}
\end{document}